# Power-dependent soliton steering in thermal nonlinear media

Fangwei Ye,[1] Yaroslav V. Kartashov,[2] Bambi Hu[1,3] and Lluis Torner[2]

[1]*Department of Physics, Centre for Nonlinear Studies, and The Beijing-Hong Kong Singapore Joint Centre for Nonlinear and Complex Systems (Hong Kong), Hong Kong Baptist University, Kowloon Tong, China*

[2]*ICFO-Institut de Ciencies Fotoniques, and Universitat Politecnica de Catalunya, Mediterranean Technology Park, 08860 Castelldefels (Barcelona), Spain*

[3]*Department of Physics, University of Houston, Houston, Texas 77204-5005, USA*

We address the existence and properties of optical solitons excited in thermal nonlinear media with a transverse refractive index gradient. We show that in such a geometry one can generate controllable switching from surface soliton propagating near the sample edges to bulk solitons. Beam steering associated to the different soliton output locations can be achieved by varying the input light intensity.

*OCIS codes: 190.5530, 190.4360, 060.1810*

Surface optical solitons are a topic of continuously renewed interest [1,2]. While most solitons are localized waves that form inside bulk media, surface solitons are localized at the interface between two different materials. Their properties substantially depend on refractive index contrast, geometry of the interface, and nonlinearities of both materials. Bulk and surface modes may feature substantially different excitation conditions. For example, surface solitons supported by truncated periodic media exhibit power thresholds for existence, in contrast to solitons in the same materials in unbounded geometries [3-7]. Surface solitons may form not only at interfaces of local materials, but also in nonlocal nonlinear media. Nonlocality substantially modifies characteristics of both bulk and surface excitations [8-17]. This is especially pronounced in media with long-range nonlocality (such as thermal media) where boundaries affect solitons propagating even in the center of the sample [11-13]. In this case boundary conditions are crucial, since they may prevent or facilitate surface soliton formation. One clear example occurs in geometries with thermally insulating interfaces, which tend to attract light [14,16].

In this Letter we address soliton formation in thermal media in a geometry that includes a gradient of linear refractive index and where one of the sample edges is kept at a fixed temperature, while other edge is thermally insulating. As expected on intuitive grounds, the linear refractive index ramp results in light deflection that counteracts the attraction towards the thermally insulating interface. Therefore, in this geometry the soliton position inside the sample depends strongly on its power, a property that may be used to generate power-controlled switching between surface and bulk solitons, hence soliton steering where the control parameter is the input light intensity.

We consider the propagation of a laser beam along the $\xi$ axis of a thermal medium occupying the region $-L/2 \leq \eta \leq +L/2$, that can be described by the system of equations for the dimensionless field amplitude $q$ and the nonlinear contribution to the refractive index $n$ that is proportional to the temperature variation, given by

$$i\frac{\partial q}{\partial \xi} = -\frac{1}{2}\frac{\partial^2 q}{\partial \eta^2} - qn + a\eta q, \quad \frac{\partial^2 n}{\partial \eta^2} = -|q|^2 \quad \text{for} \quad |\eta| \leq L/2,$$
$$i\frac{\partial q}{\partial \xi} = -\frac{1}{2}\frac{\partial^2 q}{\partial \eta^2} - qn_{\mathrm{d}} \quad \text{for} \quad |\eta| > L/2.$$
(1)

Here the transverse and longitudinal coordinate $\eta, \xi$ are scaled to the beam width and to the diffraction length, respectively; the parameter $n_{\mathrm{d}}$ describes the difference between the unperturbed refractive indices of the medium and the surrounding linear material (here we set $n_{\mathrm{d}} = 0$); the parameter $a$ describes the refractive index gradient.

In thermal media the conditions imposed at the sample boundaries affect the entire refractive index distribution. Here we consider a design where the right boundary is thermally insulating (i.e., $\partial n/\partial \eta|_{\eta=+L/2} = 0$), while the left boundary is kept at fixed temperature (i.e., $n|_{\eta=-L/2} = 0$). When laser beams enters such medium, it experiences slight absorption resulting in increase and redistribution of temperature in the entire sample. Since the right boundary is thermally insulating the heat diffusion occurs predominantly in one direction, the temperature distribution becomes asymmetric and light is deflected to the right. In the case of uniform media, the attraction to the boundary prevents formation of bulk solitons, even if the beam is launched far from the boundary.

The possibilities become richer in the presence of a preexisting linear refractive index ramp. When $a > 0$ the linear refractive index increases towards the left boundary, causing light deflection to the left. As a result, the dynamics of the beam entering such material is determined by the competition of two effects: deflection due to the linear refractive index ramp and deflection towards the thermally insulating boundary. Since the strength of attraction to boundary is determined by the power and position of the input beam, one may play with the balance among both tendencies, so that the location of beam center is given by the input power. Note that a refractive ramp may be imprinted technologically or created by using the electro-optic effect by illuminating the thermal medium with a separate modulated broad beam [18]. In all simulations we set the width of the sample to $L = 30$.

We search for soliton solutions of Eqs. (1) in the form $q = w(\eta)\exp(ib\xi)$, where $w$ is a real function and $b$ is the propagation constant. Such solitons can be characterized by their energy flow $U = \int_{-\infty}^{\infty} |q|^2 d\eta$. Our calculations confirm that in the presence of the linear refractive index ramp solitons can be located in any point inside the sample. Figures 1(a) and 1(b) show representative profiles of fundamental and dipole solitons with different propagation constant values. One can see that low-amplitude solitons reside in the vicinity of the left boundary, because the nonlinear contribution to the refractive index at low amplitudes is small and thus the effect of the linear refractive index ramp dominates, causing formation of surface waves at the left boundary. Increasing the peak intensity results in a gradual displacement of the soliton center into the bulk of the sample. For such intermediate peak amplitudes, the total refractive index $n_{\text{tot}} = n - a\eta$ features a clear maximum inside the sample [Fig. 1(c)]. For higher amplitudes, when the nonlinear contribution to the refractive index dominates, light concentrates near the right boundary of the sample.

We emphasize that the switching between the surface modes located at the left boundary, bulk modes, and surface modes located at the right boundary occurs upon continuous increase of soliton energy flow. The dependence of energy flow on propagation constant is shown in Fig. 2(a). Three different regions are clearly visible in the figure: Two regions where $U$ monotonically grows with $b$ are separated by a region where $U$ is almost constant. Solitons corresponding to the regions where $U$ increases with $b$ can be termed surface modes, because in this region soliton tails penetrate effectively into the linear medium, while solitons corresponding to the region where $U$ is almost constant can be called bulk modes. Such three regions are also visible in Fig. 2(b) showing the dependence of integral soliton center $\eta_{\text{cent}} = U^{-1} \int_{-\infty}^{\infty} \eta |q|^2 \, d\eta$ on the energy flow. As shown by the plot, the soliton center position varies most rapidly in the region where $U = U_{\text{cr}}$ is almost constant.

Light gets localized at the left boundary for $U < U_{\mathrm{cr}}$ and at the right boundary when $U > U_{\mathrm{cr}}$. Numerical integration shows that $U_{\mathrm{cr}} \approx 2a$. Such estimation can be obtained also from a qualitative analysis of the total refractive index distribution $n_{\mathrm{tot}} = n - a\eta$. The thermal contribution is given by $n(\eta) = -\int_{-L/2}^{L/2} G(\eta,\lambda)|q(\lambda)|^2 \, d\lambda$, where $G(\eta,\lambda) = -(\eta + L)$ for $\eta \leq \lambda$ and $G(\eta,\lambda) = -(\lambda + L)$ for $\eta \geq \lambda$ is the response function of thermal medium. One may roughly approximate the bulk mode profile whose width is much smaller than the width of sample as a $\delta$-function with its center located at $\eta = \eta_0$, i.e. $w(\eta) = U^{1/2}\delta(\eta - \eta_0)$. After calculation of refractive index profile one finds that $dn_{\mathrm{tot}}/d\eta = -a + U$ for $\eta < \eta_0$ and $dn_{\mathrm{tot}}/d\eta = -a$ for $\eta \geq \eta_0$. Transverse displacement may be prevented when $dn_{\mathrm{tot}}/d\eta$ is antisymmetric with respect to $\eta_0$, i.e. when $U = U_{\mathrm{cr}} = 2a$. A similar estimation may be applicable to symmetric functions $w(\eta)$ of finite width. Notice that, by its very nature, in thermal media the refractive index distribution is determined mostly by the total energy flow carried by the light beam, rather than by its particular shape. This is consistent with the fact that the rough approximation described above gives a quite accurate estimate of the critical energy flow.

The linear stability analysis that we conducted on the stationary solutions showed that all fundamental solitons are stable in the entire existence domain, irrespectively of the location of the soliton center (left boundary, bulk, or right boundary). In contrast, dipole solitons may be unstable for certain values of their propagation constant [a illustrative dependence of the real part $\delta_r$ of the perturbation growth rate $\delta = \delta_r + i\delta_i$ on $b$ for dipole solitons is shown in Fig. 2(c)].

All results discussed above are confirmed by the outcome of the direct numerical integration of Eqs. (1) with input conditions in the form of Gaussian beam whose center is initially located at $\eta_{\mathrm{in}} = 8$. Figure 3 shows three illustrative distinct propagation scenarios corresponding to different input energy flows. When $U = 0.15$ (this value is slightly smaller than the critical energy flow, which for $a = 0.08$ amounts to $U_{\mathrm{cr}} = 0.16$) the linear refractive index ramp dominates over nonlinear contribution to refractive index and the beam is deflected toward the left boundary [Fig. 3(a)]. Upon collision with the boundary it undergoes the total internal reflection and starts moving toward the center of the sample. The process repeats periodically resulting in long-living beam center oscillations. If the input energy flow coincides with the critical energy flow $U_{\mathrm{cr}} = 2a$ the balance between two competing effects is achieved and one observes stationary propagation [Fig. 3(b)]. Finally, for $U = 0.17 > U_{\text{cr}}$ one again gets near-surface oscillations, but this time near right boundary [Fig. 3(c)]. Thus, by slightly tuning the input energy flow one can vary the output soliton position in a broad range. The output beam position versus input energy flow is depicted in Fig. 2(d) for a sample of fixed length ($\xi = 300$) when the input beam is launched in the center of the sample, at $\eta_{\text{in}} = 0$. Notice that $\eta_{\text{out}} < 0$ when $U < U_{\text{cr}}$ and $\eta_{\text{out}} > 0$ for $U > U_{\text{cr}}$.

We thus conclude by stressing that adding a transverse refractive index gradient in thermal nonlinear media introduces important new soliton properties. Specifically, here we addressed a design where one edge of the material is kept at a fixed temperature while the other edge is thermally insulating. We showed that controllable soliton steering occurs in such samples, with the input light power being the control parameter.

# References with titles

# References without titles

# Figure captions

Figure 1.  Profiles of (a) fundamental solitons at $b=1$ (curve 1), 2.2 (curve 2), 7 (curve 3) and (b) dipole solitons at $b=1$ (curve 1), 2 (curve 2), and 7 (curve 3). (c) Total refractive index distributions corresponding to solitons shown in (a). In all cases $a=0.08$. Red dashed lines indicate boundaries of thermal medium.

Figure 2.  (a) Energy flow versus propagation constant and (b) position of integral soliton center versus energy flow for fundamental (black curves) and dipole (red curves) solitons. Points marked by circles in (a) and (b) correspond to solitons in Fig. 1(a). (c) Real part of perturbation growth rate versus propagation constant for dipole solitons. (d) Output beam center position versus energy flow for $\eta_{\text{in}}=0$. In all cases $a=0.08$.

Figure 3.  Propagation dynamics of solitons in thermal sample with $a=0.08$ for $U=0.15$ (a), 0.16 (b), and 0.17 (c). The initial soliton center position in all cases was $\eta_{\text{in}}=8$.

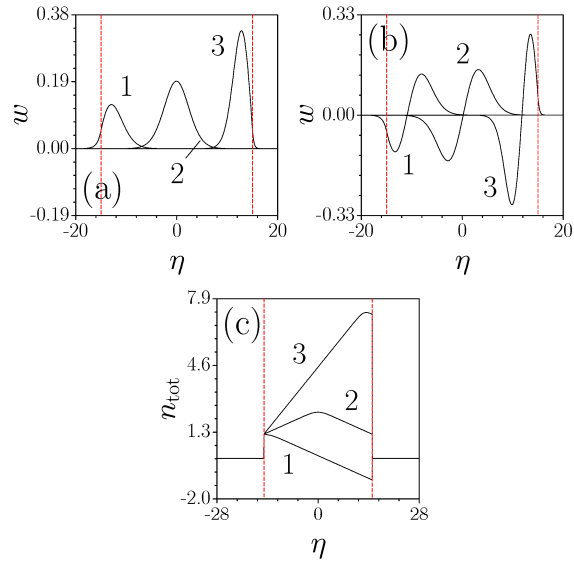

Figure 1. Profiles of (a) fundamental solitons at $b=1$ (curve 1), 2.2 (curve 2), 7 (curve 3) and (b) dipole solitons at $b=1$ (curve 1), 2 (curve 2), and 7 (curve 3). (c) Total refractive index distributions corresponding to solitons shown in (a). In all cases $a=0.08$. Red dashed lines indicate boundaries of thermal medium.

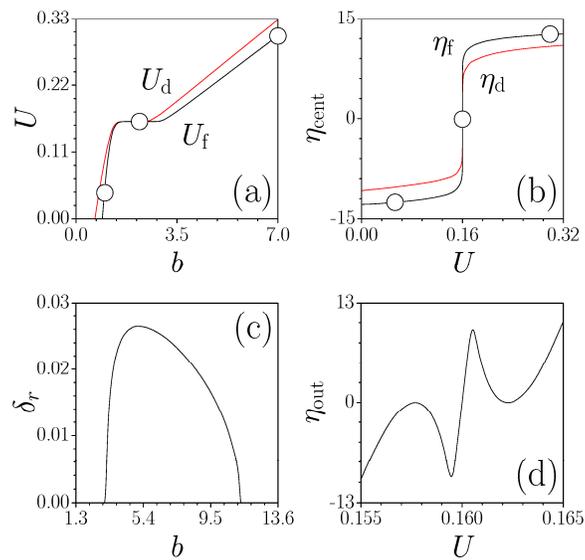

Figure 2. (a) Energy flow versus propagation constant and (b) position of integral soliton center versus energy flow for fundamental (black curves) and dipole (red curves) solitons. Points marked by circles in (a) and (b) correspond to solitons in Fig. 1(a). (c) Real part of perturbation growth rate versus propagation constant for dipole solitons. (d) Output beam center position versus energy flow for $\eta_{\rm in} = 0$. In all cases $a = 0.08$.

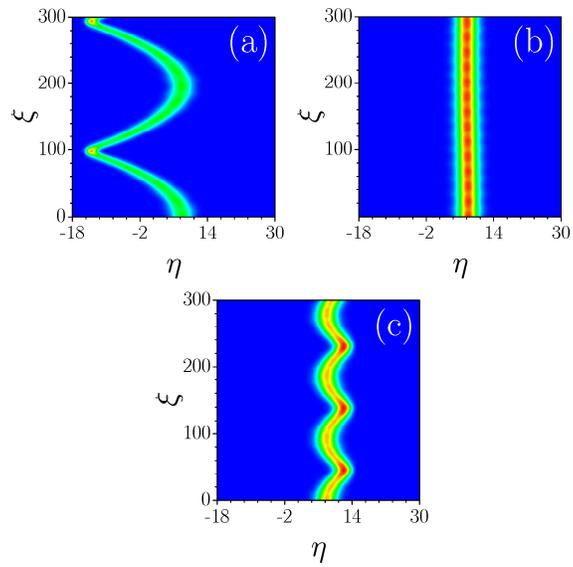

Figure 3. Propagation dynamics of solitons in thermal sample with $a = 0.08$ for $U = 0.15$ (a), 0.16 (b), and 0.17 (c). The initial soliton center position in all cases was $\eta_{\text{in}} = 8$.